\documentclass[noshowpacs,preprint,preprintnumbers,amssymb,amsmath,nofootinbib,superscriptaddress]{revtex4}
\usepackage{graphicx}
\usepackage{dcolumn}
\usepackage{bm}
\def\be{\begin{equation}} 
\def\ee{\end{equation}}
\def\bea{\begin{eqnarray}} 
\def\eea{\end{eqnarray}}
\def\line{\hbox to \hsize}    

\def \AAA{{\bf A}}

\def \CC{{\bf C}}

\def \a{{\bf a}}

\def \p{{\bf p}}

\def \n{{\bf n}}
\def \v{{\bf v}}
\def \m{{\bf m}}

\def \ket #1{{\vert #1\rangle}}
\def \bra #1{{\langle #1\vert}}

\def\eval #1#2#3{{\langle#1\vert#2\vert#3\rangle}} 

\def\1{\mbox{\bf 1}}

\newcommand{\ZZ}{{\mathbb Z}}

\newcommand{\ua}{\uparrow}
\newcommand{\da}{\downarrow}

\begin{document}

\title{Filling the Bose sea: symmetric  quantum Hall edge states  
and affine characters}

\author{Eddy Ardonne}
\email{ardonne@uiuc.edu}
\affiliation{University of Illinois, Department of Physics\\ 1110
W. Green St.\\Urbana, IL 61801 USA}

\author{Rinat Kedem}
\email{rinat@uiuc.edu}
\affiliation{University of Illinois, Department of
Mathematics\\ 1409 W. Green St.\\
Urbana, IL 61801 USA}

\author{Michael Stone}
\email{m-stone5@uiuc.edu} 
\affiliation{University of Illinois, Department of Physics\\ 1110
W. Green St.\\Urbana, IL 61801 USA}

\begin{abstract}

We explore the structure of the bosonic analogues of the $k$-clustered
``parafermion'' quantum Hall states.  We show how the many-boson wave
functions of $k$-clustered quantum Hall droplets appear naturally as
matrix elements of ladder operators in integrable representations of
the affine Lie algebra $\widehat{su}(2)_k$. Using results of Feigin
and Stoyanovsky, we count the dimensions of spaces of symmetric
polynomials with given $k$-clustering properties and show that as the
droplet size grows the partition function of its edge excitations
evolves into the character of the representation.  This confirms that
the Hilbert space of edge states coincides with the representation
space of the $\widehat{su}(2)_k$ edge-current algebra. We also show
that a spin-singlet, two-component $k$-clustered boson fluid is
similarly related to integrable representations of $\widehat{su}(3)$.
Parafermions are not necessary for these constructions.

\end{abstract}


\maketitle

\section{Introduction}

Clustered quantum Hall states were introduced by Read and
Rezayi \cite{read_parafermion} as a natural generalization
of the paired Moore-Read (or Pfaffian) state \cite{moore_read} 
that is thought to describe the FQHE phase  at  at $\nu=5/2$
\cite{greiter_wen_wilczek,morf,read_green}. The 
wave-functions of these highly-correlated electronic  states
are constructed out  correlators of   parafermion
operators \cite{zamolodchikov_fateev} combined  with a
Laughlin factor that  serves to cancel unwanted  poles 
and to maintain the overall anti-symmetry
under particle  exchange. This construction  results in
the  wave-functions  possessing  extra zeros ({\it i.e.\/} beyond  those
required by the exclusion principle) that manifest
themselves when $k+1$ electron
coordinates coincide.  These  vanishing conditions ensure that the 
clustered states are  exact zero-energy eigenstates for a
Hamiltonian with a $k$-particle repulsive interaction. As
is the case with the Pfaffian state,  the clustered phases
host quasi-particle excitations which obey non-abelian
statistics \cite{read_pfaffian_statistics,nayak_wilczek,slingerland_bais}. 

Read and Rezayi noted that  the  parafermion construction
also yields $k$-clustered boson states. It is possible
that  such  states might be realized physically in rapidly
spinning pancakes of Bose fluid \cite{cooper_wilkin_gunn}. Even  if their
creation  proves difficult, however, these
incompressible  Bose fluids are worthy of study   because
they  are in many ways simpler than their fermionic
siblings, yet they retain most of their exotic  quantum
properties. The simplification arises because  the
$k$-clustered boson wave-functions are precisely the
correlators of  the generating currents of the $SU(2)$,
level-$k$,  Wess-Zumino-Witten model --- no extra Laughlin factor being 
required. This observation suggests that  the edge-state Hilbert 
space of a large
droplet of such an incompressible Bose fluid will coincide
with the space of integrable
representations of the $\widehat{su}(2)_k$ affine Lie
algebra. It also explains why the  effective action of
for the Pfaffian Bose fluid is  a   $SU(2)$, level-$2$, Chern-Simons
theory \cite{fradkin-nayak-tsvelik-wilczek,fradkin-nayak-schoutens}.  

In order to confirm that the edge-state Hilbert
space is indeed the representation space for the current algebra, it is
useful to count the number of independent symmetric
wave-functions having  the required $k$-particle vanishing
conditions, and with a  given  polynomial degree.  If the droplet is
confined in a parabolic trap, this degree will be
proportional to the  energy of the corresponding edge state. The statistical 
partition function of the Bose droplet should then coincide with
the character for a representation of $\widehat{su}(2)_k$
\cite{lepowsky_primc}.
It turns out that this count was made by
Feigin and Stoyanovsky  \cite{feigin_stoyanovsky}
some   years before the $k$-clustered
states were introduced into physics.  
In this paper we will review Feigin
and Stoyanovsky's construction of the $k$-clustered states 
and fill in the combinatoric details of their counting
method. We
will also apply a slight modification of their construction   to
a two-component $k$-clustered phase
\cite{ardonne_schoutens,ardonne_read_rezayi_schoutens} and show that  this
leads to  the characters of the $\widehat{su}(3)_k$  current
algebra.

In section two we review the basic properties of the
integrable representations of the $\widehat{su}(2)_k$
current algebra. In section three, we will show how the
$k$-clustered  polynomials are realized as matrix elements
in a subspace  of these representations, and indicate  how
a suitable limiting procedure  (corresponding to filling
the ``Bose sea'')  should  allow us to construct the the
entire representation space. We then confirm that this
construction works  by showing that   the counting formula
for the $k$-clustered polynomials leads to  the general
level-$k$ character. We end section three by stating (without proof)
the results for the character of $\widehat{su}(3)_k$. Details of the
derivation of that result will be presented in a forthcoming paper. 
In section four we present a detailed  derivation of the counting
formula, needed to obtain the characters of $\widehat{su}(2)_k$. 

We wish to stress that there  is little original mathematics 
in the present paper.  
With the exception of our results on $\widehat{su}(3)_k$,   
nearly everything is  to be found
in \cite{feigin_stoyanovsky}. That work is, however,
written for mathematicians and is consequently rather    
inaccessible to physicists, amongst whom it deserves to be
better known. We hope that our exegesis will
be of use in this regard.

\section{The $\widehat{su}(2)_k$ Lie algebra}

To make our account self contained, and in order to establish our notation, we begin with a 
review of the  well-known properties of $\widehat{su}(2)_k$
and its representations. 

The finite $su(2)$ Lie algebra is generated by the
operators $e\equiv J_+\equiv J_1+iJ_2$, $f\equiv J_-\equiv
J_1-iJ_2$, and $h\equiv 2J_3$
with commutation relations
\be
[h,e]=2e,\quad [h,f]=-2f,\quad [e,f]=h.
\ee
The symbols $e,f,h$ are the conventional  notation in the
mathematical literature for a Lie algebra  written in
the {\it Chevalley basis\/}. $J_\pm, J_3$  are more
common in the physics literature. Mathematicians prefer to  take
$2J_3$, rather than $J_3$, as the diagonal  generator, because its
eigenvalues $\lambda\equiv 2j_3$  will be integers.
They  would  also refer to this  algebra as $sl_2$, rather
than $su(2)$, because the  ``$i$''   in the
ladder operators $J_\pm$ make it  a complexification of the real
$su(2)$ algebra. We will retain  the familiar physicist's name for the
algebra, but use the $e,f,h$ notation so
as to facilitate comparison with \cite{feigin_stoyanovsky}. 

The  infinite-dimensional $\widehat{su}(2)_k$ affine Lie
algebra \cite{kac_book,big_yellow_book} consists of linear combinations of 
operators
$e_n$, $f_n$, $h_n$, $n\in \ZZ$, together with the
central element $\hat k$. In any irreducible representation
$\hat k$ will be proportional to the identity.
Because of this, we will usually omit the 
``hat''  and simply regard $k$ as a number called the {\it
level\/} of the representation.  In all cases of interest $k$ will
be a positive integer. It is also useful to adjoin a
generator  $\hat d$ that counts the ``momentum'' $n$. The
commutation relations are then

\bea
{}&&[e_m,e_n]=[f_m,f_n]=0\nonumber\\
{}&&[e_m,\hat k]=[f_m,\hat k ]=[h_m,\hat k]=0\nonumber\\
{}&&[h_m,e_n] =2e_{m+n},\quad  [h_n,f_m] =-2f_{m+n},\nonumber\\
{}&&[e_m,f_n]= h_{m+n}+m\hat  k\delta_{n+m,0},\quad [h_m,h_n]=
2m\hat k\delta_{m+n,0},\nonumber\\
{}&&[\hat d, e_n]=ne_n,\quad [\hat d, f_n]=nf_n,\quad [\hat d,
h_n]=nh_n.
\eea

The operators $\hat k$, $\hat d$ and $h_0$ commute with each other and can
be simultaneously diagonalized. Representations of the algebra will
therefore be  spanned by vectors  
$\ket{m,\lambda, i}$ with
\be
\hat k \ket{m,\lambda,i} =k\ket{m,\lambda, i},\quad 
\hat d \ket{m,\lambda, i}= m\ket{m,\lambda,i},\quad
h_0\ket{m,\lambda,i}= \lambda\ket{m,\lambda,i}.
\ee
The eigenvalues $k$, $m$ and $\lambda$ label the {\it weights\/} of
the representation. 
The extra index $i$ in $\ket{m,\lambda,i}$ is necessary because the subspace with a given
weight will usually have dimension greater than one. We will
omit it when the weight space is one dimensional. We
also omit the label $k$ because it is the same for all
states in any given irreducible representation.

A {\it highest
weight state\/}  $\ket{{\bf v}_0}$  is a state that is annihilated by all
$e_n$, $f_n$ and $h_n$ with $n>0$, and also by $e_0$. A
{\it highest weight representation\/} is a representation containing
such a state  and such that   all other
states in the representation may be obtained by repeated 
application of the generators with $n\le 0$ to this
state. By adding a constant to $\hat d$, if necessary, we
can always take the highest weight state to obey $\hat d
\ket{{\bf v}_0}=0$.  Such a state can therefore be written
as $\ket{{\bf
v}_0}_{k,l}=\ket{0,l}$ where $l\equiv\lambda_{\rm max}$
is the largest eigenvalue of $h_0$ in a representation of the
$\{e_0,f_0,h_0\}$ finite $su(2)$ sub-algebra.

For any $n$, the generators $e_{-n}$, $f_{n}$ and $h_0-nk$
form a finite $su(2)$ sub-algebra. A representation of the
affine algebra will decompose into representations of this
finite algebra. If all such sub-representations are finite
dimensional, the original representation is said to be {\it
integrable}.  Integrable representations only occur when 
$k$ is a positive integer,  and for such a $k$   there can 
be no more than
$k+1$ inequivalent integrable representations. This restriction comes
about  because $\ket{{\bf v}_0}_{k,l}$,
being killed by $f_1$,
will be {\it lowest\/}  weight of a 
representation of the  $\{e_{-1},f_1,h_0-k\}$ 
sub-algebra. Now $(h_0-k)\ket{{\bf v}_0}_{k,l}=
(l-k)\ket{{\bf v}_0}_{k,l}$, and
$(l-k)$, being a lowest weight in  an $su(2)$ representation, must 
either be zero
or a
negative integer. Since $l$ is either zero a positive
integer, we see that $k$ must be an integer, and that  the
only possible 
values of $l$ are 
\be
l= 0,1,2,\ldots, k.
\ee 
The constraints imposed by the other $\{e_{-n},f_n,h_0-nk\}$
sub-algebras are less restrictive, and an integrable  representation
$V_{k,l}$ exists for each of these $l$ values.
There will be exactly  $(k-l)+1$ states in the $\lambda_{\rm
max}=l$ representation of
$\{e_{-1},f_1,h_0-k\}$, and so  we will have 
$(e_{-1})^{k-l+1}\ket{{\bf
v}_0}_{k,l}=0$.
This feature can be seen in  figs. 1 and 2 , where we
exhibit  the small $|m|$ parts of the weight diagrams
for the five 
integrable representations at levels $k=1,2$.

\begin{figure}
\includegraphics[width=6.0in]{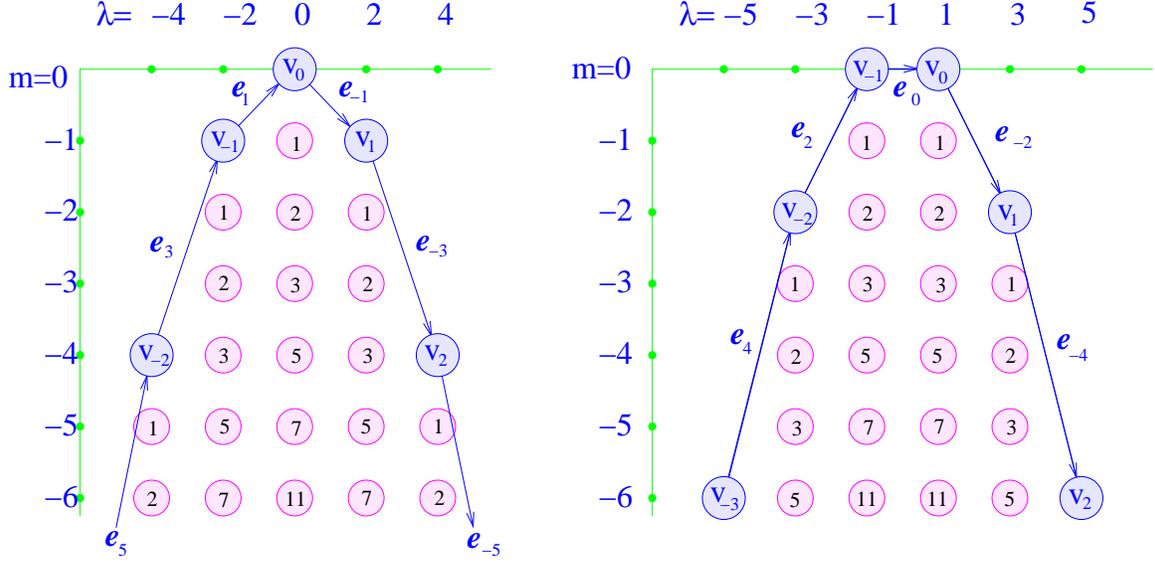}
\caption{\label{level_1} Part of the weight diagrams for for the two
integrable level $k=1$ representations $V_{1,0}$ and $V_{1,1}$. The
number in a circle is the dimension ${\rm
mult\,}_{V_{k,l}}(m,\lambda)$   of that weight
space.}
\end{figure}

\begin{figure}
\includegraphics[width=7.0in]{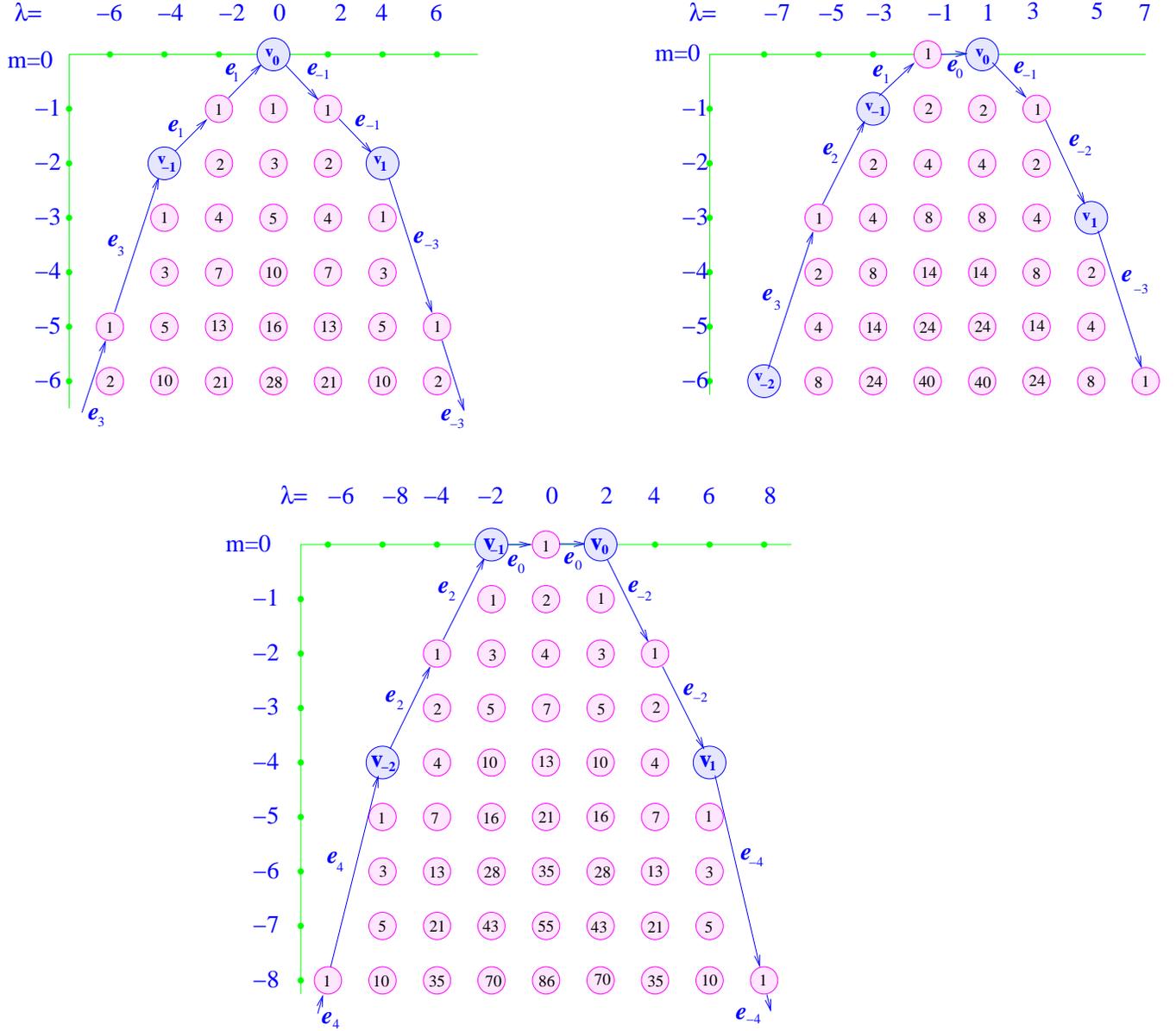}
\caption{\label{level_2} Part of the weight diagram for the level $k=2$
representations $V_{2,0}$, $V_{2,1}$ and $V_{2,2}$.}
\end{figure}

The weight diagram of a representation of a finite simple
Lie algebra is symmetric under the action of   the {\it Weyl
group\/}, which is generated by reflections in the planes
perpendicular to the root vectors. For $su(2)$, the
reflection is implemented by the action of  $S=e^{i\pi
J_1}$ on the representation. In the algebra, conjugation 
by $S$ takes $J_3\to -J_3$ and  $J_+\to J_-$.  The Weyl
symmetry is  an automorphism of the Lie algebra.  In the
affine  algebra $\widehat{su}(2)_k$, the weight diagram
is symmetric under the action of  the infinite  {\it affine Weyl group\/}. 
This group is generated by the finite $S$,
together with powers of a translation  $T$. The action of
$T$  on the algebra corresponds to conjugation by  the
element  $g(\theta)=e^{2\pi i  h\theta}$ of the associated
{ loop group\/} $\tilde{L}SU(2)$. Here $\theta$ is a
coordinate on the circle $S^1$ parameterizing the loop, and
the loop $g(\theta)$    is a  closed geodesic in  the
finite $SU(2)$ group manifold.  The affine Weyl group acts
on the weight space as   
\bea
T^n\ket{m,\lambda}&=& \ket{m-\lambda
n-kn^2,\lambda+2kn},\nonumber\\
S\ket{m,\lambda}  &=& \ket{m,-\lambda}.
\label{EQ:affine_weyl_action}
\eea
Weights related  by this transformation
will have  the same multiplicities.
We again see that integrability requires  $k$ to be an integer. If it were not, the
action of $T$ would not  preserve the $\Delta \lambda=2$
spacing of the $\lambda$
eigenvalues. 

In  figs 1 and 2 
the larger circles with labels ${{\bf v}_n}$ represent   the orbit
$\{\ket{{\bf v}_n}\equiv T^n\ket{{\bf v}_0}\/\}$ of the highest weight vector
$\ket{{\bf v}_0}$ under the action of the translation part of the
affine Weyl group. (We omit the subscripts on $\ket{{\bf
v}_ 0}$ when they are unnecessary.) The
$\ket{{\bf v}_n}$ are non-degenerate. The arrows indicate how the
states in this orbit can be obtained from one another by the
action of  suitable $e_n$'s.

We will also have cause to refer to the  Virasoro
algebra that acts  on a  representation $V_{k,l}$ 
by  
\be
L_n =\frac 1{2(k+2)} :\sum_{i+j=n}\left(e_if_j+f_ie_j +\frac
12h_ih_j\right):\,.
\ee
The normal-ordering symbols ``$:\phantom x:$'' means that 
operators with positive indices $i,j$ are placed to the right of
those with negative indices.   
Acting on $V_{k,l}$, the  $L_n$ obey 
\be
[L_n,L_m]=(n-m)L_{n+m} +\frac c{12} n(n^2-1) \delta_{n+m,0},
\ee with central charge $c= 3k/(k+2)$. 
The operator $L_0$
acts on $V_{k,l}$ as
\be
L_0= \frac{j(j+1)}{(k+2)}-\hat d.
\ee
Thus $L_0+\hat d$ is the affine analogue of  the quadratic Casimir
operator, and takes the same value on every state in the
representation $V_{k,l}$.  We also have     $[L_{-n}, e_m]=
-me_{m-n}$, together with  two
similar equations where $e_m$ is replaced by $f_m$ and
$h_m$.

\section{String Functions, Polynomials and Characters}

\subsection{String functions and characters}

The $q$-series generating function that encodes the multiplicities  in 
the column  labelled  $\lambda$ of the weight diagram of
$V_{k,l}$ 
is the  {\it string function\/} $\sigma^{k,l}_\lambda(q)$. It is
defined to be 
\be
\sigma^{k,l}_\lambda(q)= \sum_{m=0}^\infty {\rm mult\,}_{V_{k,l}}(m,
\lambda)q^{-(m-m_{0})}. 
\ee
Here $m_0$ is chosen to make the first power of $q$ in the
sum equal to $q^0=1$.

Columns that are taken
into each other  by the action of the affine Weyl group have
identical string functions. Because of this   
the two $k=1$ representations have only one distinct  string
function. For example
\be
\sigma^{1,0}_{0}(q)=1+q+2q^2+3q^3+5q^4+7q^5+11q^6+\cdots =\frac 1{(q)_\infty}.
\ee
Here the standard notation $(q)_n$ means
\be
(q)_n\equiv \prod_{m=1}^{n}(1-q^m),
\ee
and, in particular,
\be
(q)_\infty\equiv \prod_{m=1}^{\infty}(1-q^m).
\ee
The expression $Z_{\rm boson} \equiv 1/(q)_\infty$ is the 
partition function for
a free chiral boson with periodic boundary conditions, but no
winding numbers.

The representations $V_{2,0}$ $V_{2,2}$ contain two
Weyl-inequivalent columns, and  so there are two distinct
string functions 
\bea
\sigma^{2,0}_{0}(q)=1+q+3q^2 +5q^3+10q^4+16q^5+ 28q^6+\cdots &=& \frac1{(q)_\infty}
\sum_{n\,\,{\rm  even}} \frac {q^{n^2/2}}{(q)_n},  
\nonumber\\
\sigma^{2,0}_{2}(q)=1+2q+4q^2 +7q^3+13q^4+21 q^5 +35 q^6+\cdots &=& \frac
{q^{-1/2}}{(q)_\infty}\sum_{n \,\,{\rm odd}} \frac
{q^{n^2/2}}{(q)_n}.  
\eea
The other level-$2$ representation has  only one distinct
string function  
\bea 
\sigma^{2,1}_{1}(q)=1+2q+4q^2 +8q^3+14q^4+24q^5+40q^6 +\cdots &=& \frac
{1}{(q)_\infty}\sum_{n\,\, {\rm even}} \frac
{q^{n(n-1)/2}}{(q)_n},
\nonumber\\
&=&\frac
{1}{(q)_\infty}\sum_{n\,\,{\rm  odd}} \frac
{q^{n(n-1)/2}}{(q)_n}
\nonumber\\
&=&\frac 1 {(q)_\infty}\prod_{n=1}^{\infty}(1+q^n).
\label{EQ:string3}
\eea
The three equalities in (\ref{EQ:string3}) 
can be extracted  from the more general
$q$-series \cite{andrews}
\be
\sum_{n=0}^{\infty} \frac{q^{n(n-1)/2}x^n}{(q)_n}=
(1+x)(1+xq)(1+xq^2) \cdots.
\ee 
The product 
\be
Z^{+}_{\rm Majorana} \equiv \prod_{n=1}^{\infty}(1+q^n)
\ee
appearing in the last line of (\ref{EQ:string3}) 
is the partition function for an arbitrary number ({\it
i.e.\/} chemical potential zero) of  free chiral Majorana
fermions with
periodic boundary conditions. Combining the sum over the
even and odd integers that occurs in the $l=0,2$
representations gives  
\be
 \sum_{n=0}^\infty \frac {q^{n^2/2}}{(q)_n}=Z^{-}_{\rm
 Majorana}\equiv 
\prod_{n=1}^{\infty} (1+q^{n+1/2}),
\ee
which is the partition function for an arbitrary number of free chiral 
Majorana fermions with
anti-periodic boundary conditions. The sum over even integers
captures that part of the partition function with an even
number of Majorana particles and the sum over the odd
integers the part with  an odd number of particles%
\footnote{The ubiquity of  Majorana fermions at $k=2$ is connected
with the three primary WZW fields $\phi^{j=1}_m$,
$m=-1,0,1$,  having scaling
dimension $1/2$, and so being identifiable as linear combinations of  the triplet of
Majorana fermion fields that generate the $k=2$ current
algebra \cite{fradkin-nayak-tsvelik-wilczek}.}.

The string functions are  assembled into  the characters
\be \label{EQ:stringchar}
{\rm ch}_{V_{k,l}}(q,x)= \sum_{m,\lambda} {\rm
mult\,}_{V_{k,l}}(m,\lambda) q^{-m} x^\lambda.
\ee
Thus, for example,
\bea
{\rm ch}_{V_{1,0}}(q,x)&=& \frac
1{(q)_\infty}\sum_{n=-\infty}^{\infty} q^{n^2}x^{2n},
\nonumber\\
{\rm ch}_{V_{1,1}}(q,x)&=& \frac
1{(q)_\infty}\sum_{n=-\infty}^{\infty}
q^{n(n+1)}x^{2n+1}.
\label{EQ:level_one_characters}
\eea

\subsection{Symmetric polynomials}

To obtain the many-boson wave functions, we begin by 
defining  the field
\be
e(z)=\sum_{n=-\infty}^{\infty} e_{n} z^{-n-1}.
\ee
This is not quite the $e(z)$ appearing in  \cite{feigin_stoyanovsky},
but is the customary   expansion of a 
field of conformal dimension one. With our  definition, for
example, 
\bea
{}[L_{-1}, e(z)]
&=&\partial_z e(z)\nonumber\\
{}[L_0, e(z)]&=& e(z) +z\partial_z e(z).
\eea

Because $e_n\ket{{\bf v}_0}=0$ for all
$n\ge 0$, the equation 
$(e_{-1})^{k+1-l}\ket{{\bf v}_0}=0$ can be written as
\be
\left.(e(z))^{k+1-l}\ket{{\bf v}_0}\right|_{z=0}=0.
\ee
Also  the equation  $\big(e(z)\big)^{k+1}=0$ holds as an operator identity in
any  integrable representation of level $k$. That this is
true for the $V_{k,0}$ representation follows immediately from
$\big(e(0)\big)^{k+1}\ket{{\bf v}_0}=0$ by using the infinitesimal
translation property of  $L_{-1}$, coupled with that fact
that  $L_{-1}\ket{{\bf v}_0}_{k,0}=0$.
We can show, by manipulations with the  
primary fields of the associated WZW model, that   
$\big(e(z)\big)^{k+1}=0$ remains true in the other integrable
representations.       

Now consider an integrable  representation $V=V_{k,l}$, and in it  
the sub-space $W$ spanned by a states of the form 
$e_{-i_1}e_{-i_2}\ldots e_{-i_m}\ket{{\bf v}_0}$. We
construct the matrix elements
\be
F_{\bf v}(z_1,z_2,\ldots,z_p)\equiv 
\eval{{\bf v}}{e(z_1)e(z_2)\ldots e(z_p)}{{\bf v}_0}
\ee
where $\ket{\bf v}$ lies in one of the weight spaces in $V$.  There  are  at most  a finite number
of products of $e_{-n}$ that can take us from $\ket{{\bf
v}_0}$ to $\ket{{\bf v}}$, and so
$F_{\bf v}(z_1,z_2,\ldots,z_p)$
will be a { polynomial\/} in the variables $z_i$.
Because  the $e_n$  commute with each other,  
$F_{\bf v}(z_1,z_2,\ldots,z_p)$ is a {\it symmetric\/} polynomial.
If $\ket{{\bf v}}=\ket{m,\lambda,i}$, this
polynomial will be homogeneous of degree ${\rm deg\,}(F_{\bf v})$,
with 
\bea
-m&=&{\rm deg\,}(F_{\bf v})+p,
\nonumber\\ \label{polketrel}
\lambda &=& l+2p.
\eea

The equations $\big(e(z)\big)^{k+1}=0$ and $(e_{-1})^{k+1-l}\ket{{\bf v}_0}=0$
force  the  polynomials to possess the following properties:

\begin{itemize}

\item[i)] For all ${\bf v}$, the polynomial 
$F_{\bf v}(z_1,z_2,\ldots,z_p)$ must vanish when more than
$k$ of the $z_i$ coincide. (It may also   vanish when  $k$
or fewer points coincide.)

\item[ii)] For all $\bf v$, the polynomial 
$F_{\bf v}(z_1,z_2,\ldots,z_p)$ must  vanish if $k+1-l$
of the $z_i$ become zero. (Again, it   may also vanish 
when fewer than this number of $z_i$ gather at $z=0$.)

\end{itemize}

\noindent  
The  $F_{\bf v}$ 
thus  have the $k$-cluster vanishing
properties, and, in addition, have extra zeros located at the
origin that correspond to the insertion there of $k-l$ 
quasi-holes. When multiplied by a suitable
gaussian factor $\exp(-\Omega \sum_i |z_i|^2/4)$ and with
$z=x+iy$, they become  the  many-body wave functions of the
two-dimensional, $k$-clustered, Bose
gas phase.

There is a one-to-one
correspondence between states in the sub-space $W$ and the
those in the space 
\be
{\mathcal F}=\bigoplus_{p=0}^\infty {\mathcal F}_p,
\ee
where ${\mathcal F}_p$ is the space of polynomials  $F_{\bf
v}$ in  $p$ variables. 
To see why this is so, we  need first to realize
that the $F_{\bf v}$ are  really labelled by the  bra vector
$\bra{{\bf v}}$, which  is an element of the
dual space $W^*$. We are therefore claiming  that the
linearly independent $k$-clustered symmetric polynomials form a basis for
$W^*$. If there
were no dependencies between between the $e_i$ then $W^*$
would be the entire space of symmetric polynomials---this
being the usual correspondence between the first- and
second-quantized version of a bosonic Fock space. There are,
however, linear relations 
between   the vectors $e_{-i_1}e_{-i_2}\ldots e_{-i_p}\ket{{\bf v}_0}$
due to the vanishing of the  Fourier components of
$\left(e(z)\right)^{k+1}$. Thus 
\be
S_m=\sum_{i_1+i_2+\ldots +i_{k+1}=m} e_{i_1}e_{i_2}\ldots
e_{i_{k+1}}=0,
\ee
where we can restrict the sum to $e_i$ with $i<0$. There
are also relations arising from  $(e_{-1})^{k-l+1}$ being
zero when acting on the vacuum. The $S_m$'s, together with 
$(e_{-1})^{k-l+1}$,  generate an  {\it ideal\/} $\mathcal
I$ in the free commutative algebra $U(e)$ consisting of
linear combinations of  monomials in the $e_i$, $i<0$.  The
space $W$ can  be identified with the quotient algebra
$U(e)/{\mathcal I}$. In order to be uniquely defined as an
element of $W^*$, any  $\bra{{\bf v}}\in W^*$  must give
zero when paired with a state in the space  ${\mathcal
I}\ket{\bf v}_0$. This, however, is precisely what the
vanishing properties of the polynomials enforce. The
linearly independent  polynomials do therefore form a basis
for $W^*$.  

We now observe that, although $W$ and $W^*$ are both
infinite dimensional, the individual weight spaces are
finite dimensional. There is therefore a one-to-one
correspondence between $\ket{m,\lambda,i}\in W$ and the
dual basis $\bra{m,\lambda,i}\in W^*$.   The dimensions of
the space of $k$-clustered polynomials of appropriate 
degree, and in an appropriate number of variables, do
therefore coincide with the dimensions of the weight spaces
in $W$.

Not every state in  $V$ can be obtained as 
$e_{-i_1}e_{-i_2}\ldots e_{-i_p}\ket{{\bf v}_0}$. Since
each application of an $e_{-n}$ moves us one step to the
right and $n$ steps down, we cannot reach any weight to the 
left of the central column in the weight diagram. Furthermore, we can reach
weights close to the central column {\it via\/}   fewer distinct
products $e_{-i_1}e_{-i_2}\ldots e_{-i_p}$ than the
dimension of the weight space.  Because of this paucity of
paths, even if all the
$e_{-i_1}e_{-i_2}\ldots e_{-i_p}\ket{{\bf v}_0}$ were linearly
independent, we would only be able to obtain  a restricted
class of  
states in the weight space. If, however, we look at weights
in  columns far to the  right in  the weight diagram,  the
number of distinct products leading to a given  weight  grows
rapidly, whilst the  dimensions of the weight spaces near
the head  of any such  column remain small. It is
plausible, therefore (and we will see that it is   true),
that we can obtain all  states in such a weight space by
applying suitable products of $e_{-n}$'s to $\ket{{\bf
v}_0}$.  Thus, by counting the number of symmetric
polynomials  with the given vanishing properties, we can
obtain the early terms in the string functions. Then, by
taking suitable limits, we can obtain {\it all\/} the terms in the
string functions.  

\subsection{Filling the sea}

To illustrate these ideas, consider the two $k=1$ representations. The
general $p$-variable symmetric 
polynomial satisfying the $l=0$ vanishing conditions is 
\be
F(z_1,z_2,\ldots,z_p)=S(z_1,z_2,\ldots,z_p)\prod_{i<j}(z_i-z_j)^2,
\ee
where $S(z_1,z_2,\ldots,z_p)$ is a general symmetric polynomial in  
$p$ variables. The number of general symmetric  polynomials
of degree $n$ in $p$ variables 
is given by  the coefficient of $q^n$ in $1/(q)_p$, and the degree of
the remaining factor is $p(p-1)$. Using $-m= {\rm
deg}(F)+p$ and $\lambda=2p$, we find that  the
contribution of the $e_{-i_1}\ldots e_{-_p}\ket{{\bf
v}_0}\in W$   to the 
${\rm ch}_{V_{1,0}}$
character is
\be
{\rm ch}_{W}(q,x)=\sum_{p=0}^{\infty}\frac 1{(q)_p}
 q^{p^2} x^{2p}.
\ee  
Comparing this with the full character
(\ref{EQ:level_one_characters}), and observing that
the first $p$ terms in $1/(q)_p$ and $1/(q)_\infty$
coincide, we see that the
first $p$ weights in each column are correctly counted.

For the case $k=1, l=1$, the general polynomial is
\be
F(z_1,z_2,\ldots,z_p)=S(z_1,z_2,\ldots,z_p)\prod_{i,j}(z_i-z_j)^2
\prod_{i=1}^pz_i. 
\ee
The product factors have  degree $p^2$. 
The contribution of the $e_{-i_1}\ldots e_{-i_p}\ket{{\bf v}_0}$
states is therefore
\be
{\rm ch}_{W}(q,x)=
\sum_{p=0}^{\infty}\frac 1{(q)_p} q^{p(p+1)} x^{2p+1}.
\ee
Again comparison with (\ref{EQ:level_one_characters}) shows
that  the first $p$ states in each column are correctly
counted. 

To get the entire representation we therefore look at the 
states near the heads of the  columns surrounding $\ket{{\bf
v}_N}$.
We map these weights back to the
neighbourhood of $\ket{{\bf v}_0}$  by using the Weyl
translation $T^{-N}$ and so find all the states  near the
highest-weight  ``vacuum''. We then send $N$ to
infinity, and so find all the states in an arbitrarily large
neighbourhood  $\ket{{\bf v}_0}$. 

The state $\ket{{\bf v}_N}$ corresponds to an undeformed
droplet  containing $kN$ bosons.  In mapping it back {\it
via\/} the affine Weyl group to the
``vacuum'' state $\ket{{\bf v}_0}=T^{-N}\ket{{\bf v}_N} $ 
we also map the original highest-weight state  to $\ket{{\bf v}_{-N}}$,
which becomes the new no-particle state.  The relation
\be
\ket{{\bf v}_0}= e_0^le_{1}^{k-l}\ket{{\bf v}_{-1}},
\ee
which is illustrated by the arrows in the weight diagrams,
can be generalized to
\be
\ket{{\bf v}_0}= (e_0^le_{1}^{k-l})( e_{2}^le_{3}^{k-l})\ldots
(e_{2N-2}^le_{2N-1}^{k-l})\ket{{\bf v}_{-N}}.
\ee
This process of adding $Nk$ particles to the no-particle
state corresponds to building up the droplet by filling
the ``Bose sea'' .  That we only get a full representation of
the current algebra  in the limit of a large droplet is
familiar from the case of the $\nu=1$ Hall effect
\cite{stone_schur}.

\subsection{Level-$k$ characters}

We now write down the $q$-series generating function that
counts  the general level-$k$ polynomials.  Because it is
rather lengthy, we  defer  the  proof of this counting
formula to  the next section. 

When there are  $p$ variables $z_i$, the generating function  takes the form
of   a sum over the partitions of $p$ into   parts, each of
which is no  larger than $k$.   These $k$-restricted
partitions are most conveniently pictured  as Young
diagrams whose shape is described  by  the integers
$m_1,\ldots, m_k$. Here $m_\alpha$ is the number of rows
containing $\alpha$ boxes.
These definitions are  illustrated in  fig.~3.

\begin{figure}
\includegraphics[width=3.0in]{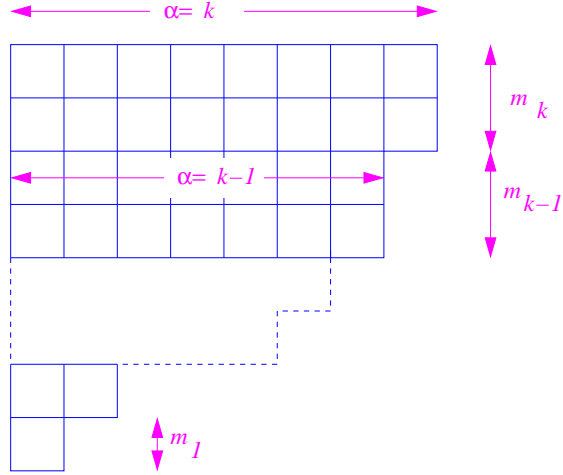}
\caption{\label{partition_defs} Young diagram  representing
a $k$-restricted partition of $p$.}   
\end{figure} 

For the spaces ${\mathcal F}_p(d)$  of  $k$-clustered
symmetric polynomials of degree $d$ in $p$ variables, Feigin
and Stoyanovsky show that  \cite{feigin_stoyanovsky}  
\be
\sum_d {\rm mult\,}(p,d)\,q^{d} = q^{-p}
\sum_{\substack{ k-{\rm restricted}\\ {\rm partitions}\\ {\rm of}\; p}}
\frac{q^{{\bf
m}^t {\bf M}{\bf m} +{\bf d}^t{\bf m}}}{(q)_{m_1}
(q)_{m_2}\ldots (q)_{m_k}}.
\label{EQ:level_k_counting}
\ee
Here the $m_\alpha$ are the positive integers describing the
partition of $p$,
and so they are constrained by   
\be
m_1+2m_2+\cdots + km_k =p.
\ee
The 
matrix ${\bf M}$ has entries $M_{\alpha\beta}={\rm
min}(\alpha,\beta)$. The vectors ${\bf m}$ and ${\bf d}$
are defined by
\bea
{\bf m}&=&(m_1,m_2,\ldots,m_k)^t,\\
{\bf d}&=&(0,\ldots,0, 1,2,\ldots, l)^t,
\eea
the non-zero entries in ${\bf d}$ beginning at $\alpha
=k-l+1$.

From (\ref{EQ:level_k_counting}) we obtain the  contribution to the
character of the states in the restricted subspace $W$. We
do this    by setting $-m=d+p$, $\lambda =2p+l$, and summing over $p$. Thus
\be
{\rm ch}_{W_{k,l}}(q,x)=\sum_{p=0}^{\infty}x^{2p+l}\left\{\sum_{m_1+2m_2+\ldots km_k=p}
 \frac{ q^{{\bf
m}^t{\bf  M}{\bf m} +{\bf d}^t{\bf m}}}{(q)_{m_1}
(q)_{m_2}\ldots (q)_{m_k}}\right\}.
\ee   
The quadratic form in the exponent can be diagonalized by
setting
\bea 
N_1&=& m_1+m_2+\cdots +m_k,\nonumber\\
N_2&=& m_2+\cdots +m_k,\nonumber\\ 
   &\vdots&\nonumber\\
N_k&=& m_k, \label{EQ:transform}
\eea
whence
\be
{\bf m}^t{\bf M}{\bf m} =N_1^2+N_2^2+\cdots +N_k^2.
\ee
In terms of the $N_\alpha$, we have
\be
{\rm ch}_{W_{k,l}}(q,x)=
\sum_{p=0}^{\infty}x^{2p+l}\left\{\sum_{N_1+\cdots+N_k=p}
 \frac{ q^{{N_1^2+N_2^2+\cdots N_k^2+ N_{k-l+1}+N_{k-l+2} +\cdots
 N_{k}}}}{(q)_{N_1-N_2}
(q)_{N_2-N_3}\ldots (q)_{N_k}}\right\}.
\ee Here the  $N_\alpha$ are restricted by $N_1\ge N_2\ge
\cdots\ge N_k\ge 0$.  

The diagonalized version of the character 
makes it easier to express the  partial character   that
counts  the states in 
the pulled-back space $T^{-N}W$. 
Applying $T^{-N}$ to the weights in 
\be
{\rm
ch}_{W_{k,l}}(q,x) = \sum_{\lambda,m}{\rm
mult\,}_W(\lambda,m)x^\lambda q^{-m},
\ee
we obtain 
\be
{\rm
ch}_{T^{-N}W_{k,l}}(q,x) = \sum_{\lambda,m}{\rm
mult\,}_W(\lambda,m)x^{\lambda-2kN} q^{-m-\lambda N+ kN^2}.
\ee
To apply   this transformation, we  observe  that
\be
\lambda \equiv 2p+l = 2(N_1+N_2+\cdots+N_k) +l,
\ee
and that there are  $l$ terms in the linear part of
the exponent 
$$
N_{k-l+1}+N_{k-l+2} +\cdots
 N_{k}.
$$ 
This suggests that we complete the squares in the quadratic form,
and then   change  variables   
$N_\alpha\to N_\alpha-N$. We end up with   
obtain
\be
{\rm ch}_{T^{-N}W_{k,l}}(q,x)=\sum_{p=0}^\infty
x^{2(p-Nk)+l}\left\{\sum_{N_1+\cdots+N_k=p-Nk}
 \frac{q^{{N_1^2+N_2^2+\cdots N_k^2+ N_{k-l+1}+N_{k-l+2} +\cdots
 N_{k}}}}{(q)_{N_1-N_2}
(q)_{N_2-N_3}\ldots (q)_{N_{k-1}-N_k}(q)_{N_k+N}}\right\},
\ee
where  $N_1\ge N_2\ge
\cdots\ge N_k\ge -N$.
It is now straightforward to  take  the limit $N\to
{\infty}$ and recover   
the character for the complete
representation
\be
{\rm
ch}_{V_{k,l}}(q,x)=\lim_{N\to \infty}{\rm
ch}_{T^{-N}W_{k,l}}(q,x).
\ee
We find 
\be \label{EQ:character_su2kl}
{\rm ch}_{V_{k,l}}(q,x) = \frac 1{(q)_\infty} \sum_{N_1\ge
N_2\ge \ldots\ge N_k}
 \frac{ x^{2N_1+2N_2\cdots +2N_k+l}
q^{{N_1^2+N_2^2+\cdots N_k^2+ N_{k-l+1}+N_{k-l+2} +\cdots
 N_{k}}}}{(q)_{N_1-N_2}
(q)_{N_2-N_3}\ldots (q)_{N_{k-1}-N_k}},
\ee
where $N_1\ge N_2\ge
\cdots\ge N_k\in {\bf Z}$. This is indeed the full character
of the $V_{k,l}$  representation
\cite{lepowsky_primc}, and
so we have verified  the assertion that  all states are obtainable by
application of products of  $e_{-n}$ to the no-particle
state $\ket{{\bf v}_{-\infty}}_{k,l}$.

The effect of filling the Bose
sea is to  make the 
width $k$ part of the Young
diagram extend indefinitely upwards. This infinite tail   
corresponds to an undisturbed core of $k$-clustered
bosons, whilst  the  $\alpha<k$ part represents  a halo  of partially 
disrupted clusters composing the edge excitations.    

We close this subsection with a rather appealing form of the character
(\ref{EQ:character_su2kl}), which will also establish the notation
used in the next subsection. 
This form is obtained by using the inverse of the transformation
(\ref{EQ:transform}), with the result
\be
{\rm ch}_{V_{k,l}}(q,x)=\frac{1}{(q)_\infty} 
\sum_{\substack{m_1,m_2,\ldots,m_{k-1}\geq0\\m_k \in \ZZ}}
x^{2p+l} \frac{ q^{\frac{1}{2} {\bf m}^t{\bf 2 M}{\bf m} +{\bf d}^t{\bf m}}}
{(q)_{m_1} (q)_{m_2}\ldots (q)_{m_{k-1}}} \ ,
\ee
where $p=\sum_{\alpha=1}^k \alpha m_\alpha$. In the context of the
clustered quantum Hall states, the matrix ${\bf 2M}$ first appeared in
\cite{ardonne_bouwknegt_schoutens}.
After performing a final transformation
\be
m_k = \frac{p-\sum_{\alpha=1}^{k-1} \alpha m_\alpha}{k} \ ,
\ee
the character takes the form
\be
{\rm ch}_{V_{k,l}}(q,x) =  
\sum_{p \in \ZZ} x^{\AAA_1 p + l} q^{\frac{p \AAA_1 p}{2k} +
\frac{l p}{k}}
\left(
\frac{1}{(q)_\infty}
\sideset{}{'}\sum_{m_1,\ldots,m_{k-1}}
\frac{q^{\frac{1}{2}{\bf m}^t \AAA_1 \AAA_{k-1}^{-1} {\bf m} -
(\AAA_{k-1}^{-1} {\bf m})_{k-l}}}
{(q)_{m_1}(q)_{m_2}\ldots(q)_{m_{k-1}}} \right) \ ,
\ee
where $\AAA_{k-1}^{-1}$ is the inverse Cartan matrix of
$su(k)$. Explicitly,
$(\AAA_{k-1}^{-1})_{\alpha\beta}=\min(\alpha,\beta) -
\frac{\alpha\beta}{k}$ and
$(\AAA_{k-1})_{\alpha\beta}= 2 \delta_{\alpha,\beta} -
\delta_{|\alpha-\beta|,1}$, where $\alpha,\beta=1,\ldots,k-1$.
We used the convention that $(\AAA_{k-1}^{-1} {\bf m})_{k-l}$ is zero for
$l=0$ and $l=k$.

Note that $\AAA_1=2$. The prime indicates the constraint
$\sum_{\alpha=1}^{k-1} \alpha m_\alpha = p \mod k$. The appealing
feature of this form of the character is that it's explicitly
expressed in terms of the string functions, which are proportional to
the expression in parenthesis, 
c.f.\ eq.\ \eqref{EQ:stringchar}. In addition, the
character is determined by $k,l$, the Cartan matrix of $su(2)$, and
the Cartan matrix of $su(k+1)$.

\subsection{The case $\widehat{su}(3)_k$.} 
In \cite{ardonne_schoutens}, spin-singlet analogues of the
spin-polarized Read-Rezayi states were proposed (see
\cite{ardonne_read_rezayi_schoutens} for more details). These states
have the same clustering property, namely that the wave function
vanishes if any $k+1$ electron coordinates coincide. We can exploit 
the ladder operator  strategy to construct  characters
of $\widehat{su}(3)_k$  in a manner analogous to that of 
the previous
sections for $\widehat{su}(2)_k$. However, because of the
presence of the two spin components, the situation is slightly more
involved. We will be brief in this section, and refer to a forthcoming
paper \cite{forthcoming} for details.

We will here consider only the vacuum representation $V_k$ of 
$\widehat{su}(3)_k$ (see \cite{forthcoming} for arbitrary
representations). In the vacuum representation, consider
the subspace $W$, that  is spanned by states of the form
$e_{\alpha_1,-i_1}e_{\alpha_1,-i_2}\ldots e_{\alpha_1,-i_{m_1}}
f_{\alpha_2,-i_1}f_{\alpha_2,-i_2}\ldots f_{\alpha_2,-i_{m_2}}
\ket{{\bf v}_0}$. This is {\it not\/} the subspace $W$
exploited 
by Feigin and Stoyanovsky \cite{feigin_stoyanovsky}. They 
consider the sub-space
generated by the action of $e_{\alpha_1,-i}$ and
$e_{\alpha_2,-j}$
on $\ket{{\bf v}_0}$. This  may seem more natural from
the viewpoint of Lie theory --- after all $\alpha_1$ and $\alpha_2$
are  the simple roots of the algebra --- but our space has the advantage
that $e_{\alpha_1, -i}$ and $ f_{\alpha_2,-j}$ commute.

As in  $\widehat{su}(2)_k$ case, we
construct the functions
\be
F_{\bf
v}(z^\ua_1,z^\ua_2,\ldots,z^\ua_{p_1};z^\da_1,z^\da_2,\ldots,z^\da_{p_2})
\equiv  \eval{{\bf v}}{e_{\alpha_1}(z^\ua_1)e_{\alpha_1}(z^\ua_2)\ldots
e_{\alpha_1}(z^\ua_{p_1})
f_{\alpha_2}(z^\da_1)f_{\alpha_2}(z^\da_2)\ldots
f_{\alpha_2}(z^\da_{p_2})}{{\bf v}_0} \ .
\ee
With our choice of commuting operators, 
these  functions are symmetric polynomials in the variables $z^\ua_i$, 
and separately in the variables $z^\da_i$.
The polynomials $F_{\bf v}$ can therefore be interpreted as the
coordinate part of wave functions 
of  quantum Hall states, in which the spin-up (spin-down)
electrons are located at $z_i^\ua$ $(z_j^\da)$ respectively.

In \cite{feigin_stoyanovsky} the objects corresponding to $F_\v$ are
rational functions (which therefore can not be interpreted as wave
functions for the spin-singlet quantum Hall systems), and the presence
of poles complicates the derivation of the character.

For similar reasons to those adduced  the $\widehat{su}(2)_k$ case, we have
a set of  relations
$(e_{\alpha_1}(z))^{k+1-a}(f_{\alpha_2}(z))^a=0$,
$a=0,1,\ldots,k+1$, which are valid  in any integrable
representation.
For the special case of the vacuum  representation, we have
additional relations
$(e_{\alpha_1,-1})^{k+1} \ket{\v_0}=0$ and $f_{\alpha_2,0}
\ket{\v_0}=0$. From these relations, it follows that $F_\v$ must be
zero if any $k+1$  variables 
 coincide (and it  may vanish when fewer
variables coincide).

Before we can state the character of the subspace $W$, we first have
to introduce a  labelling of  the weight spaces. As was the case for
$\widehat{su}(2)_k$, the representations of $\widehat{su}(3)_k$ are
spanned by vectors $\ket{\v} = \ket{m,\lambda,i}$. In this case,
however, $\lambda\equiv \lambda_1 \omega_1+\lambda_2\omega_2 $ is a 
weight of $su(3)$. Here $\omega_1$ and $\omega_2$ are the
fundamental weights (the highest weights of  the $3$ and $\bar 3$
representations) and  $\lambda_1$ and $\lambda_2$ are
integers.
\begin{figure}
\includegraphics[width=2in]{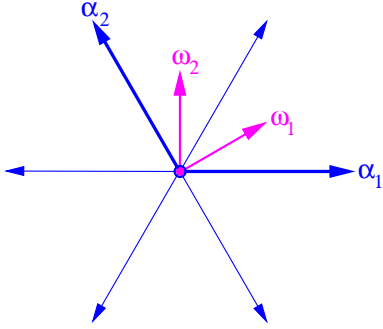}
\caption{The roots and weights of $su(3)$. The vectors
$\alpha_1$ and $\alpha_2$ are the simple roots, and
$\omega_1$ and $\omega_2$ are the fundamental weights.}
\label{su3_roots}
\end{figure}
For $\ket{\v} = \ket{m,\lambda,i}$, the polynomial
$F_{\bf
v}(z^\ua_1,z^\ua_2,\ldots,z^\ua_{p_1};z^\da_1,z^\da_2,\ldots,z^\da_{p_2})$
is of homogeneous degree $\deg (F_\v)$, with (compare with the
analogous relations for $\widehat{su}(2)_k$ in \eqref{polketrel})
\bea
-m & = &\deg (F_\v) + p_1 + p_2 \nonumber \\
\lambda & = & (\lambda_1,\lambda_2) = (2p_1+p_2,-(p_1+2p_2)) \ .
\eea

The character for the subspace $W$ now takes the following form
\be \label{su3kw}
{\rm ch}_{W_{k,0,0}} (q,x_1,x_2) = 
\sum_{\substack{m_1,\ldots,m_{k}\geq 0 \\
n_1,\ldots,n_{k}\geq 0}}
\frac{
 (x_1)^{2p_1+ p_2} (x_2)^{-(p_1+2p_2)}
 q^{\frac{1}{2}(\m^t {\bf 2M} \m + \n^t {\bf 2M} \n + \m^t {\bf 2 O} \n)}
}
{\prod_{\alpha=1}^{k} (q)_{m_\alpha} (q)_{n_\alpha}} \ ,
\ee
where the matrix ${\bf O}$ has entries
$O_{\alpha\beta} = \max(0,\alpha+\beta-k)$, $p_1=\sum_{i=1}^k i m_i$
and $p_2=\sum_{i=1}^k i n_i$. The matrix corresponding to
bilinear form in \eqref{su3kw} first appeared in the context of the
non-abelian quantum Hall states in \cite{ardonne_bouwknegt_schoutens}. 

To find the character of the vacuum  representation, we have to fill
the sea by applying
the appropriate Weyl translation $T^{-n}_{\alpha_1-\alpha_2}$, in a
manner similar to the $\widehat{su}(2)_k$ case (c.f.\ eq.\ 
\eqref{EQ:affine_weyl_action}). We then take the limit $n\rightarrow\infty$.
The resulting character can be written in the following form
\be \label{su3kchar}
{\rm ch}_{V_{k,0,0}} (q,x_1,x_2) = \frac{1}{(q)_\infty^2}
\sum_{p_1,p_2\in \ZZ} (x_1)^{2p_1+ p_2} (x_2)^{-(p_1+2p_2)}
q^{\frac{1}{2k} \p^t \bar{\AAA}_2 \p}
\sideset{}{'}
\sum_{\substack{m_1,\ldots,m_{k-1}\geq0\\n_1,\ldots,n_{k-1}\geq0}}
\frac{q^{\frac{1}{2} \a^t \AAA_2 \otimes \AAA_{k-1}^{-1} \a}}
{\prod_{i=1}^{k-1} (q)_{m_i} (q)_{n_i}} \ ,
\ee
where $\a^t = (m_1,\ldots,m_{k-1},n_1,\ldots,n_{k-1})$,
$F_1=\sum_{\alpha=1}^{k-1} \alpha m_\alpha$,
$F_2=\sum_{\beta =1}^{k-1} \beta n_\beta$ and
$\p^t =(p_1,p_2)$. $\bar{\AAA}_2$ is obtained from the Cartan matrix of
$su(3)$ by changing the sign of the off-diagonal elements.
The prime indicates two constraints,
$\sum_{\alpha=1}^{k-1} \alpha m_\alpha =p_1 \mod k$ and
$\sum_{\beta=1}^{k-1} \beta n_\beta = - p_2 \mod k$.

We have restricted the discussion to the vacuum representation,
because the other representations involve some additional
complications, which will be described in \cite{forthcoming}.

The character formula \eqref{su3kchar} for the vacuum
representation of
$\widehat{su}(3)_k$ can be generalized to representations $V_{k,\lambda}$
with $\lambda = l w_i$ of any simply-laced affine Lie algebra
$\widehat{g}_k$. To do this, we
need to replace $\AAA_2$ by the Cartan matrix $\CC_r$ of the
corresponding finite Lie algebra $g$, whose rank we denote by $r$. In
addition, we need to change the dependence on $x_j$. 

In the end, the character of the vacuum representation
takes the following form 
\be \label{sllachar}
{\rm ch} V_{\widehat{g}_k} (q,\{x_i\})= \frac{1}{(q)_\infty^r}
\sum_{p_1,\ldots,p_r \in \ZZ}
\prod_{i=1}^{r} (x_i)^{(\CC_r \p)_i}
q^{\frac{1}{2k} \p^t \CC_r \p}
\sideset{}{'}
\sum_{\substack{m_1^{(1)},\ldots,m_{k-1}^{(1)}\geq0\\
\vdots\\
m_1^{(r)},\ldots,m_{k-1}^{(r)}\geq0}}
\frac{q^{\frac{1}{2} \m^t \CC_r \otimes \AAA_{k-1}^{-1} \m}}
{\prod_{j=1}^{r} \prod_{\alpha=1}^{k-1} (q)_{m_\alpha^{(j)}}} \ ,
\ee
where $\p^t=(p_1,\ldots,p_r)$ and
$\m^t=(m_1^{(1)},\ldots,m_{k-1}^{(1)};\cdots;m_1^{(r)},\ldots,m_{k-1}^{(r)})$,
while the prime indicates the constraints
$\sum_{\alpha=1}^{k-1} \alpha m_\alpha^{(j)}= p_j \mod k$, for $j=1,\ldots,r$.
The character for the basic representation of
$\widehat{su}(3)_k$, eq. \eqref{su3kchar}, can be brought into the
form of eq. \eqref{sllachar} by changing the sign of the summation
variable $p_2$.  Note that the character formula eq.\
\eqref{sllachar} appeared earlier, in references \cite{kirillov} and
\cite{georgiev}. 
The strategy of our proof, which will appear in a
future publication \cite{forthcoming}, is the same as the approach
used in this paper for $\widehat{su}(2)_k$ and is rather different from
the combinatorical approach of \cite{georgiev}.

\section{Counting the polynomials}

In this section we expand on the sketch provided in  
\cite{feigin_stoyanovsky} and explain  how to  derive
(\ref{EQ:level_k_counting}). This requires us to compute  the
dimensions of the spaces ${\mathcal F}_p(d)$ of  
degree-$d$, symmetric polynomials in $p$ variables that
satisfy the $k$-cluster vanishing conditions and are
additionally  zero when $z_1=z_2=\ldots =z_{k-l+1}=0$. We
will do this by introducing a {\it filtration\/} associated
with $k$-restricted partitions  of $p$ on the total space 
\be
{\mathcal
F}_p= \bigoplus_{d=0}^{\infty} {\mathcal F}_p(d)
\ee
of such functions. Our discussion is a simplified version
of that appearing in \cite{rinat}.

In this section we will follow the convention that
partitions are labelled  by  greek letters
such as $\lambda$ and $\mu$. 
The context  should prevent  confusion with  
the eigenvalues of $h_0$.   We  represent a partition
$\lambda$  by a 
Young diagram whose shape is parameterized by positive
integers $m_\alpha$, as described in the previous section  and
illustrated in figure 3. We
order the partitions of $p$ lexicographically ---{\it i.e.\/} we  
say that $\lambda>\mu$, if, as we read down from the top of
the Young diagram and come to the first  row that differs between
the two partitions, the row in $\lambda$ is the longer.  This
is  a { total order relation}: any two
partitions $\lambda$ and $\mu$  are either identical
or one is strictly greater than the other.

To define the filtration we write  the variables $z_i$ into
the boxes of the partition $\lambda$ in any order. We will relabel them
so that  $z^{(\alpha)}_{ij}$ is the variable  in the
$j$'th column of the $i$-th row of length $\alpha$. 
Given a partition we now define the {\it evaluation map\/}
$\varphi_\lambda$ that acts  on a function in ${\mathcal
F}_p$ 
by setting the variables in each row equal to a common value
$z_i^{(\alpha)}$. The result is a polynomial in $\sum
m_\alpha$ variables that is symmetric under the interchange
of variables with the same value of $\alpha$---{\it
i.e.\/} deriving from rows of the same length.
We claim that the image  ${\mathcal H}_\lambda$ of
${\mathcal F}_p$ under the action of $\varphi_\lambda$ 
is the space of linear combinations of   polynomials  of the form
\be
 H(\{z_i^{(\alpha)}\}) =h(\{z_i^{(\alpha)}\}) H_\lambda
 \label{EQ:functions_in_H1},
\ee
 where  $h(\{z_i^{(\alpha)}\})$ is any function symmetric under the interchange
of variables with the same value of $\alpha$, and 
\be 
 H_\lambda=\prod_{\alpha= k-l+1} ^k \prod_{i=1}^{m_\alpha}
 (z_i^{(\alpha)})^{\alpha-k+l} \prod_{(\alpha,
 i)>(\alpha',i')} (z_i^{(\alpha)}-
 z_{i'}^{(\alpha')})^{2M_{\alpha\alpha'}}.
 \label{EQ:functions_in_H2}
 \ee
 Here $M_{\alpha\beta}= {\rm min\,}(\alpha,\beta)$,   
and 
 we are
thinking of the row index $i$  increasing downward as is customary
when writing matrices, so---perhaps perversely---we order
the indices so that $(\alpha,i)>(\alpha',i')$ if
$\alpha>\alpha'$, or, if $\alpha=\alpha'$, then $i<i'$.   
Now set
\be
\Gamma_\lambda =\bigcap_{\mu>\lambda} {\rm
ker\,}\varphi_\lambda,\quad  \Gamma'_\lambda =\bigcap_{\mu\ge \lambda} {\rm
ker\,}\varphi_\lambda.
\ee
Thus $\Gamma_\lambda$ is the space of functions that
annihilated by every evaluation map for $\mu>\lambda$.
Clearly  $\Gamma_\lambda \subset \Gamma_\mu$ if
$\lambda>\mu$.  This nested set of subspaces of ${\mathcal F}$ is the filtration
that we require.  

Observe that   ${\Gamma'}_\lambda \subset
\Gamma_\lambda$, 
and ${\Gamma'}_{(1^p)}=0$ because  $\varphi_{(1^p)}$
conflates no variables,  and is therefore an isomorphism. 
We can therefore define the graded space
\be
{\rm Gr} \Gamma=\bigoplus_\lambda {\rm Gr}_\lambda \Gamma
\ee
where ${\rm Gr}_\lambda \Gamma
=\Gamma_\lambda/{\Gamma'}_\lambda$ and the sum is over
partitions of $p$. Our strategy is   to prove that 
\be
\varphi_\lambda:{\rm Gr}_\lambda  \Gamma\to {\mathcal
H}_\lambda
\ee
is an isomorphism of graded vector spaces. We must therefore
show that  this map is well defined, and that it is both
injective and surjective. Because of this isomorphism, we can use the
polynomials in ${\mathcal H}_\lambda$ rather than the polynomials in
${\mathcal F}_p$ to calculate the character, which makes the problem
tractable. 

We begin by showing that the map  is injective and  well-defined. To do this
first observe that the image of ${\Gamma'}_\lambda$ under
the action $\varphi_\lambda$ is automatically zero as a
consequence of the
definition of  ${\Gamma'}_\lambda$. There is therefore no
problem in defining the  action of
$\varphi_\lambda$ on the quotient space
${\rm Gr}_\lambda \Gamma= \Gamma_\lambda/{\Gamma'}_\lambda$.
Furthermore, by definition,  the difference of any two functions in
$\Gamma_\lambda$ that map
down to the same function in ${\mathcal H}_\lambda$ lies in
${\Gamma'}_\lambda$. The map  is therefore injective.

To complete the demonstration  that the map is well-defined, we must 
show our characterization of  the space ${\mathcal H}_\lambda$  is
correct  in that its elements are indeed of the form
claimed in (\ref{EQ:functions_in_H1}) and
(\ref{EQ:functions_in_H2}). To show that $f\in {\mathcal
H}_\lambda$
implies that $\varphi_\lambda(f)$ has a zero of at least $2\,{\rm
min\,} (\alpha,\beta)$ when
$z_i^{(\alpha)}=z_{i'}^{(\beta)}$,
it is sufficient to consider the dependence of $f$ on the
two set of variables $\{z_{ij}^{(\alpha)}\}_{j=1}^\alpha$
and $\{z_{i'j}^{(\beta)}\}_{j=1}^\beta$, with $\alpha\ge
\beta$. We can carry out the evaluation map in two steps:
 $\varphi_\lambda=\varphi^1\circ\varphi^2$. Here $\varphi^1$
 consists of conflating all sets of variables except the 
 $\{z_{i'j}^{(\beta)}\}_{j=1}^\beta$, and $\varphi^2$
 consists of setting $z_{i'1}^{(\beta)}= \cdots= 
 z_{i'\beta}^{(\beta)}=z_{i'}^{(\beta)}$. Let
 \be
 f_1(z_i^{(\alpha)};
 z_{i'1}^{(\beta)},\ldots,z_{i'\beta}^{(\beta)})
 =\varphi^1[f(\{z_i\})]
 \ee
 Now, by definition $f(z_i)$ is annihilated  by all  $\varphi_\mu$
 with $\mu>\lambda$. Therefore 
 \be
 \left.f_1(z_i^{(\alpha)};z_{i'1}^{(\beta)},\ldots, 
z_{i'\beta}^{(\beta)})\right|_{z_{i'j}^{(\beta)}=z_i^{(\alpha)}}=0\quad
j=1,\ldots,\beta,
\label{EQ:f_1_zero}
\ee
because this corresponds to an evaluation at some partition
larger than $\lambda$. Therefore,
 \be
 f_1(z_i^{(\alpha)};
 z_{i'1}^{(\beta)},\ldots,z_{i'\beta}^{(\beta)})=
 \prod_{j=1}^\beta
 (z_i^{(\alpha)}-z_{i'j}^{(\beta)})\tilde f_1(z_i^{(\alpha)};
 z_{i'1}^{(\beta)},\ldots,z_{i'\beta}^{(\beta)}).
\label{EQ:firstorder_zero}
 \ee
 
Now $f_1(z_i^{(\alpha)};
 z_{i'1}^{(\beta)},\ldots,z_{i'\beta}^{(\beta)})$ was
 obtained from a symmetric function, and so, for each $j$,
 \be
 \left.\frac{\partial f_1}{\partial z_i^{(\alpha)}}
 \right|_{z_{i'j}^{(\beta)}=z_i^{(\alpha)} }  
= \alpha \left.\frac{\partial f_1}{\partial z_{i'j}^{(\beta)}}
 \right|_{z_{i'j}^{(\beta)}=z_i^{(\alpha)}}.
 \label{EQ:derivative1}
\ee
However (\ref{EQ:firstorder_zero}) tells us that, again for
each $j$, 
\be 
\left.\frac{\partial f_1}{\partial z_i^{(\alpha)}}
 \right|_{z_{i'j}^{(\beta)}=z_i^{(\alpha)}}  
= - \left.\frac{\partial f_1}{\partial z_{i'j}^{(\beta)}}
\right|_{z_{i'j}^{(\beta)}=z_i^{(\alpha)}}=\left.{
\sideset{}{'}\prod_{j'=1}^\beta}
 (z_i^{(\alpha)}-z_{i'j'}^{(\beta)})
 \tilde f_1\right|_{z_{i'j}^{(\beta)}=z_i^{(\alpha)}},
\label{EQ:derivative2}
\ee
the prime on the product meaning that the term with $j'=j$ is
to be omitted.
The only way to reconcile (\ref{EQ:derivative1}) with
(\ref{EQ:derivative2}) is for 
$ \tilde f_1|_{z_{i'j}^{(\beta)}=z_i^{(\alpha)}}$ to be zero.
Thus the  zero at $z_{i'j}^{(\beta)}=z_i^{(\alpha)}$
is  at least  a double zero:  
\be
 f_1(z_i^{(\alpha)};
 z_{i'1}^{(\beta)},\ldots,z_{i'\beta}^{(\beta)})=
 \prod_{j=1}^\beta
 (z_i^{(\alpha)}-z_{i'j}^{(\beta)})^2\tilde f_2(z_i^{(\alpha)};
 z_{i'1}^{(\beta)},\ldots,z_{i'\beta}^{(\beta)}).
\label{EQ:f_1_zero2}
\ee
We now evaluate the right-hand-side of (\ref{EQ:f_1_zero2})
at
$z_{i'1}^{(\beta)}=\ldots=z_{i'\beta}^{(\beta)}=z_{i'}^{(\beta)}$
and, recalling the condition that $\alpha\ge\beta$, we have
\be
\varphi_\lambda[f(\{z_i\})] = (z_i^{(\alpha)}-
z_{i'}^{(\beta)})^{2\,{\rm min\,}(\alpha,\beta)}\tilde f.
\label{EQ:A_matrix}
\ee

Now we must establish that the image of $\varphi_\lambda$
acting on $f\in {\mathcal F}$ has a zero of order at least ${\rm
max\,}(0, \alpha-k+l)$ whenever $z_i^{(\alpha)}=0$. To see
that this is so, consider the dependence of $f$ on the
variables $z_1,\ldots z_\alpha$, with $\alpha>k-l$, 
which are set equal to each other under the mapping
$\varphi_\lambda$. We carry our the map by setting the
variables equal to zero consecutively. The function 
\be
g(z_{k-l+1},\ldots, z_\alpha;\ldots)
=f|_{z_1=\ldots=z_{k-l}=0},
\ee  
has a factor $\prod_{i=k-l+1}^\alpha z_i$ because it is zero
if any of the remaining variables are set to equal to zero.
Therefore if any of the $z_i^{\alpha}$ in the evaluation
mapping are set to zero, the image has a zero of at least
degree $\alpha-k+l$.

The last, and hardest, task is to establish surjectivity. In other
words, to show that there is at least one pre-image function in
${\mathcal F}_p$ for   every   function of  the form
(\ref{EQ:functions_in_H1},\ref{EQ:functions_in_H2}) that we claim
constitutes the image ${\mathcal H}_\lambda$.

\begin{figure}
\includegraphics[width=3.0in]{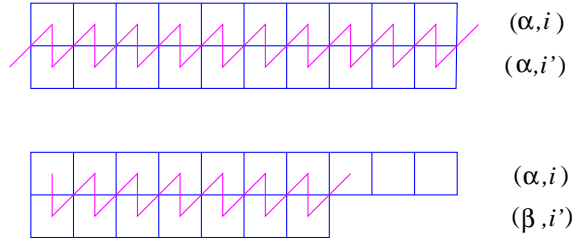}
\caption{\label{FIG:wiring} The pattern on pairwise products
appearing in equation (\ref{EQ:function_f}). The top
figure is for the case $\alpha=\beta$ and the second for
$\alpha>\beta$. Each line joining the centers of a box
corresponds to a factor $(z_i-z_j)$.}   
\end{figure}

Let $\lambda$ be a fixed partition and consider a function
$F(\{z_i\})$ of the form
\be
F(\{z_i\}) = {\rm Sym\/} f(\{z_i\})
\ee
where the symmetrization is over the variables $z_i$ and
\be
f(\{z_i\})
=\tilde h(\{z_i\})\prod_{(\alpha,i)>(\beta,i')}(z_{ij}^{(\alpha)}-z_{i'j}^{(\beta)})
(z_{i,j+1}^{(\alpha)}-z_{i'j}^{(\beta)})\prod_{\alpha>k-l}\prod_{
i,j>k-l} z_{ij}^{(\alpha)} ,
\label{EQ:function_f}
\ee
where $\tilde h$ is a polynomial, and $z_{i,\alpha+1}^{(\alpha)} \equiv z_{i,1}^{(\alpha)}$.
(The pattern of pairwise products is illustrated in figure
\ref{FIG:wiring}.)
We claim that the function $F$ is an element of
$\Gamma_\lambda$. 

We first observe
that if any $k+1$ of the $z_i$ are set equal,  the
factor $\prod  (z_{ij}^{(\alpha)}-z_{i'j}^{(\beta)})$ is
zero: Consider the Young diagram corresponding to $\lambda$.
There are most $k$ columns. Since we are equating $k+1$
variables, at least two of them are forced to be in the same
column and so give  to a zero factor.  
It is easy to check that $F$ satisfies the vanishing
conditions  $z_1=\ldots =z_{k-l+1}=0$. It is  therefore an
element of $\mathcal F_p$.

We next show that $F$ is in the kernel of $\varphi_\lambda$
for any $\mu>\lambda$: We arrange the variables
$z_{ij}^{(\alpha)}$ in the boxes of the Young diagram
corresponding to $\mu$ and recall that variables in the same
row are set equal to each other under the evaluation map.
Thus no two variables from the same column in $\lambda$ may
appear in the same row in $\mu$ if the prefactor 
$\prod  (z_{ij}^{(\alpha)}-z_{i'j}^{(\beta)})$ is to be
non-zero. Since $\mu>\lambda$, this is impossible: Let the
lengths of the rows in the partitions be denoted by
$\lambda_i$ and $\mu_i$, with the index $i$ increasing
downwards. Suppose that $\mu_i=\lambda_i$ for $i=1,\ldots,j$
and $\mu_j>\lambda_j$. All the variables from the rows
$\lambda_i=k$ must appear in the rows of $\mu$ with the same
length, similarly for rows of length $k-1$, and so on. But
in row $j$ of $\mu$ there can be at most $\lambda_j$
variables from row $j$ of $\lambda$, and at least one other
variable must come from row $j'$ with $j'>j$. But
$\lambda_{j'}\le \lambda_j$, and hence this variable belongs
to the same column of $\lambda$ as some variable in row $j$.
Therefore under the evaluation mapping the image of $F$ is zero.

Now we turn to the factor $\tilde h$ which is to give rise
to the symmetric function $h$ in (\ref{EQ:functions_in_H1}).
We can take as a basis for such functions the symmetrized
monomials
\be
h=\prod_{\alpha} \sum_{\sigma\in S_{m_\alpha}} \sigma
\left(\prod(z_i^{(\alpha)})^{\lambda^{(\alpha)}_i}\right)
\label{EQ:function_h}
\ee
Here $\lambda^{(\alpha)}$ is a partition with no more that
$m_\alpha$ rows,  these rows being of length
$\lambda_i^{(\alpha)}$. The symmetrization is over each set
of variables $(z_1^{(\alpha)},\ldots
z_{m_\alpha}^{(\alpha)})$ for fixed $\alpha$. 
Let
\be
\tilde h(\{z_i\}) =\prod_{\alpha=1}^k\left[\prod_{i=1}^{m_\alpha}
(z_{i1}^{(\alpha)})^{\lambda_i^{(\alpha)}}
\right],
\ee
be the function appearing in (\ref{EQ:function_f}). We claim that
$\varphi_\lambda(F)$ is a scalar multiple of the function
$H$ in (\ref{EQ:functions_in_H1}) with $h$ as in
(\ref{EQ:function_h}).

To see that this claim is correct consider the sum over the
symmetric group $S_p$ 
\be
\sum_{\sigma\in S_p} f(\sigma(\{z_i\})).
\ee
Suppose that for some $\sigma$, we have
$\sigma(z_{ab}^{(\beta)})=z_{cd}^{(\gamma)}$ with
$(\gamma,c)<(\beta,a)$, and that $(\beta,a)$ is the largest
row for which this true. This means that all rows above
$(\beta,a)$ undergo only a permutation within the row.  
Suppose that the prefactor 
\be
\varphi_\lambda\circ\sigma \left(
\prod_{(\alpha,i)>(\alpha',i')}(z_{ij}^{(\alpha)}-z_{i'j}^{(\alpha')})
(z_{i,j+1}^{(\alpha)}-z_{i'j}^{(\alpha')})\right) 
\ee
is to be non-zero. Then $z_{cd}^{(\gamma)}$ cannot be in  a
column directly below or to the left of the
permutation image of any other element from row $(\beta,a)$
---but this means that at least one other element from row
$(\beta,a)$ should be mapped to a row below $(\beta,a)$. If
it is mapped to the $(\gamma, c)$ it can appear in any
column other than $d$. If it mapped to any other row, it can
appear in any other column  than $d$ and an adjacent column
(to the right or left depending on whether it is above or
below $(\gamma,c)$.) Now we repeat this argument for this
new element, concluding that at least one more element of
row $(\beta,a)$ is mapped to a lower row, and so forth,
until eventually we find that all elements are permuted
to a row below $(\beta,a)$. 
If the elements are permuted to the same row, they can be placed
adjacent columns. Elements which are permuted to different rows can
not be placed in adjacent columns, this being due to the factor
linking adjacent columns in the prefactor.  
There are at most $\beta$ columns in $\lambda$ in rows below
$(\beta,a)$,
and hence the elements must all appear in the same row, which is
therefore of length $\beta$. Thus all the variables in rows of length $\beta$
are mapped to another row of length $\beta$, for the same
reason. As a result, he only permutations which give a non-zero
contribution to $\varphi_\lambda\circ\sigma$ are those that
which permute variables within each row, or those that
permute rows of equal length.
Under the evaluation map, the
former contribute equal terms to the sum, while row
interchanges correspond to the symmetrization over
$z_i^{(\alpha)}$ in (\ref{EQ:function_h}).

We have now completed the proof that the map
$\varphi_\lambda:{\rm Gr}_\lambda \Gamma \to {\mathcal H}_\lambda$ is
an isomorphism. This map is clearly degree preserving, and so
it only remains to count the number of polynomials in each
space $H_\lambda$.  

Referring back to  (\ref{EQ:functions_in_H2}) we see that the
minimal
degree of a polynomial associated with a 
partition $l$ receives a contribution from the second product of 
\hbox{$2\,
m_\alpha m_\beta\, {\rm
min\,}(\alpha,\beta)$}  for each pair $\alpha>\beta$, and
\hbox{$
\alpha\, m_\alpha(m_\alpha-1)$}  for each $\alpha$. The first
product makes a contribution \hbox{$\sum_\alpha m_\alpha {\rm
min\,}(0,\alpha-k+l)$.} The total minimal polynomial degree
is therefore
\be
\sum_{\alpha\beta} m_\alpha M_{\alpha\beta}\, m_\beta
-\sum_\alpha \alpha\, m_\alpha +\sum_\alpha\, m_\alpha  {\rm
min\,}(0,\alpha-k+l).
\ee 
Using $\sum_\alpha \alpha m_\alpha =p$, the total number of
variables, this becomes 
\be 
{\bf m}^t{\bf M}{\bf m}+ {\bf d}^t{\bf m} -p,
\ee
where ${\bf m}=(m_1,\ldots m_k)^t$ and ${\bf d}$ has entries
$d_i ={\rm
min\,}(0,\alpha-k+l)$. The generating functions for the
number of 
symmetric polynomials $h(\{z_i^{(\alpha)}\})$ of degree $d$ are
$\left(\prod_\alpha (q)_{m_\alpha}\right)^{-1}$. Putting the
parts together, we therefore have
\be 
\sum {\rm mult\,}(p,d)\,q^{d} = q^{-p}
\sum_{\substack{ k-{\rm restricted}\\ {\rm partitions}\\ {\rm of}\; p}}
\frac{q^{{\bf
m}^t {\bf M}{\bf m} +{\bf d}^t{\bf m}}}{(q)_{m_1}
(q)_{m_2}\ldots (q)_{m_k}},
\ee
as claimed in (\ref{EQ:level_k_counting}).

\section{Conclusion}

We have shown that  a planar droplet of bosons in a strongly correlated
Read-Rezayi $k$-clustered phase \cite{read_parafermion}  
will have low energy  excitations that can be
identified with states in  representations of the
$\widehat{su}(2)_k$ current algebra. This simple picture
applies provided that  the  energies of these excited states are  
sufficiently low  that only the edge of the droplet is
perturbed,  a central core being left intact.  This picture
provides a physical interpretation for the observation
\cite{feigin_stoyanovsky} that the  space of $k$-clustered
symmetric polynomials in an arbitrary number of variables is in
one-to-one correspondence with states in the integrable
representations of the Lie algebra $\widehat{su}(2)_k$.
We also presented a generalization of these results to a two component
spin-singlet system, whose excitations are classified by the
$\widehat{su}(3)_k$ algebra.

\section{Acknowledgments}

This work was supported by the National Science Foundation
under grant NSF-DMR-01-32990. MS and EA thank Eduardo
Fradkin for many useful conversations, and RK thanks 
Boris Feigin and Tetsuji Miwa for discussions about  counting methods.

\end{document}